\documentclass{article}
\usepackage[margin=1in, includefoot]{geometry}
\usepackage{amsmath}
\usepackage{graphicx}
\usepackage{float}
\usepackage{media9}
\usepackage{hyperref}
\usepackage{multicol}
\usepackage[labelformat=empty]{caption}
\usepackage[backend=biber, style=nature, citestyle=numeric-comp, sorting=none]{biblatex}
\providecommand{\keywords}[1]{\textit{Keywords:} #1}

\newcommand\sbullet[1][.5]{\mathbin{\vcenter{\hbox{\scalebox{#1}{$\bullet$}}}}}
\begin{document}
\begin{center}
\LARGE{A framework for parsing heritable information}\\
\vspace{10pt}
\large{Antony M Jose\footnote[1]{correspondence: email - amjose@umd.edu, mail - Rm 2136, Bioscience Research Building, 4066 Campus Drive, University of Maryland, College Park, MD - 20742, USA.}}\\
\small{\textit{Department of Cell Biology and Molecular Genetics,}}\\
\small{\textit{University of Maryland,}}\\
\small{\textit{College Park, MD 20742, USA}}\\
\vspace{5pt}
\begin{abstract}
Living systems transmit heritable information using the replicating gene sequences and the cycling regulators assembled around gene sequences. Here I develop a framework for heredity and development that includes the cycling regulators parsed in terms of what an organism can sense about itself and its environment by defining entities, their sensors, and the sensed properties. Entities include small molecules (ATP, ions, metabolites, etc.), macromolecules (individual proteins, RNAs, polysaccharides, etc.), and assemblies of molecules. While concentration may be the only relevant property measured by sensors for small molecules, multiple properties that include concentration, sequence, conformation, and modification may all be measured for macromolecules and assemblies. Each configuration of these entities and sensors that is recreated in successive generations in a given environment thus specifies a potentially vast amount of information driving complex development in each generation. This Entity-Sensor-Property framework explains how sensors limit the number of distinguishable states, how distinct molecular configurations can be functionally equivalent, and how regulation of sensors prevents detection of some perturbations. Overall, this framework is a useful guide for understanding how life evolves and how the storage of information has itself evolved with complexity since before the origin of life.
\end{abstract}
\keywords{systems biology, information theory, homeostasis, transgenerational inheritance, synthetic biology}
\end{center}
\begin{multicols}{2}
\vspace{5pt}
\noindent\textbf{Introduction}
\vspace{5pt}

Analyses of living systems from molecular to population scales have revealed information storage and processing across multiple scales as key attributes of life \cite{Tkacik2016}. The need to understand the behavior of a basic unit of life - a single cell - in terms of an integrated framework for information handling has been previously articulated \cite{Tanaka1984, Nurse2008, Danchin2009, Brenner2010}, but is yet to be developed. A single cell is often the bottleneck stage that separates successive generations, making it the minimal space for storing all heritable information (see supplemental text for variations on the single-cell bottleneck). Such information in molecules are part of the `nature' of organisms and do not include information transmitted when parents train progeny, which can be considered as `nurture'. Cells and more complex living systems can change their information content by learning through interactions with their environment. However, their ability to transmit any such learned information from one generation to the next is limited by the available storage in the bottleneck stage and potentially other system constraints (e.g. inability of learned information to cross generational boundaries)\cite{Jose2018}. To appreciate these limits, we need to consider the total amount of information that could be encoded using \textit{all} molecules in the bottleneck stage. Such joint consideration of all heritable information that is transmissible using molecules will inform how complexity grows over evolutionary time, what constitutes nature versus nurture, and how to synthesize new living systems.

To facilitate discussion of all heritable information, I begin by defining key terms introduced in an earlier article \cite{Jose2018}: stores of information, stored information, and cell code. \textit{Stores}(n.) of information refer to molecules or arrangements of molecules that hold information. This information can be transferred to other molecules or arrangements and the original store can be degraded or modified after such transfer. Therefore, molecules and arrangements of molecules can have \textit{stored}(v.) information. \textit{Cell code}(n.) refers to the heritable information that \textit{encodes} the development of organisms from a bottleneck stage, which is minimally a single \textit{cell}. Similar  development in successive generations in a given environment presumably relies on similar cell codes assembled during bottleneck stages (see supplemental text for more on assembly of cell codes).

The information in a cell code can be conceptually separated into two distinct forms \cite{Jose2018}. One is the genome sequence, where information is stored in a linear sequence of bases, and the other is the recurring arrangement, where information is stored in the concentrations, configurations, and interactions of molecules in bottleneck stages (see supplemental text on cell code assembly). While the information content in this arrangement and the extent to which it is recreated is currently not easily quantified, it is clear that heredity relies on information that is held in multiple stores and transmitted across generations. This communication of heritable information through the development of an organism from one generation to the next has been likened to the transmission of messages through a communication channel from sender to receiver (e.g. refs. \cite{Battail2010, Jose2018}). Just as `the fundamental problem of communication is that of reproducing at one point either exactly or approximately a message selected at another point' \cite{Shannon1948}, the fundamental problem of heredity is that of reproducing at one bottleneck stage either exactly or approximately a cell code selected at the preceding bottleneck stage.

Each choice of molecules and their arrangement in bottleneck stages collectively stores heritable information and forms a message transmitted across generations. The average information content in a message chosen from among \textit{N} possible messages is given by the Shannon entropy, $H = \sum_{i=1}^{N} -p_i \sbullet log_2(p_i)$, where $p_i$ is the probability of the $i^{th}$ message \cite{Shannon1948, Cover2006}. If the probability of selecting each message is equal, this expression simplifies to give the maximal information a message can carry, $H = \sum_{i=1}^{N} -1/N \sbullet log_2(1/N) = log_2N$ bits. Therefore, to determine the maximal heritable information in a living system, we need to enumerate all distinguishable states of its bottleneck stage (i.e., $N$). This exercise will provide a starting point for the joint analysis of all heritable information that needs to be transmitted across generational boundaries for the reproduction of living systems.

\vspace{5pt}

\noindent\textbf{Static and dynamic storage of information}

\vspace{5pt}
The logical requirements for self-replication have been explored in two-dimensional universes called cellular automata using abstract `machines' \cite{Sipper1998}\footnote[2]{Individual units in these machines are called `cells', but are referred to as `parts' in this article to avoid confusion with biological cells.}. Of particular relevance are self-replicating machines that use the same store of information in two distinct ways: (1) as instructions whose interpretation leads to the construction of an identical copy of the machine, and (2) as data to be copied without interpretation and placed in the copied machine.
\begin{figure}[H]
  \includegraphics[width=\linewidth]{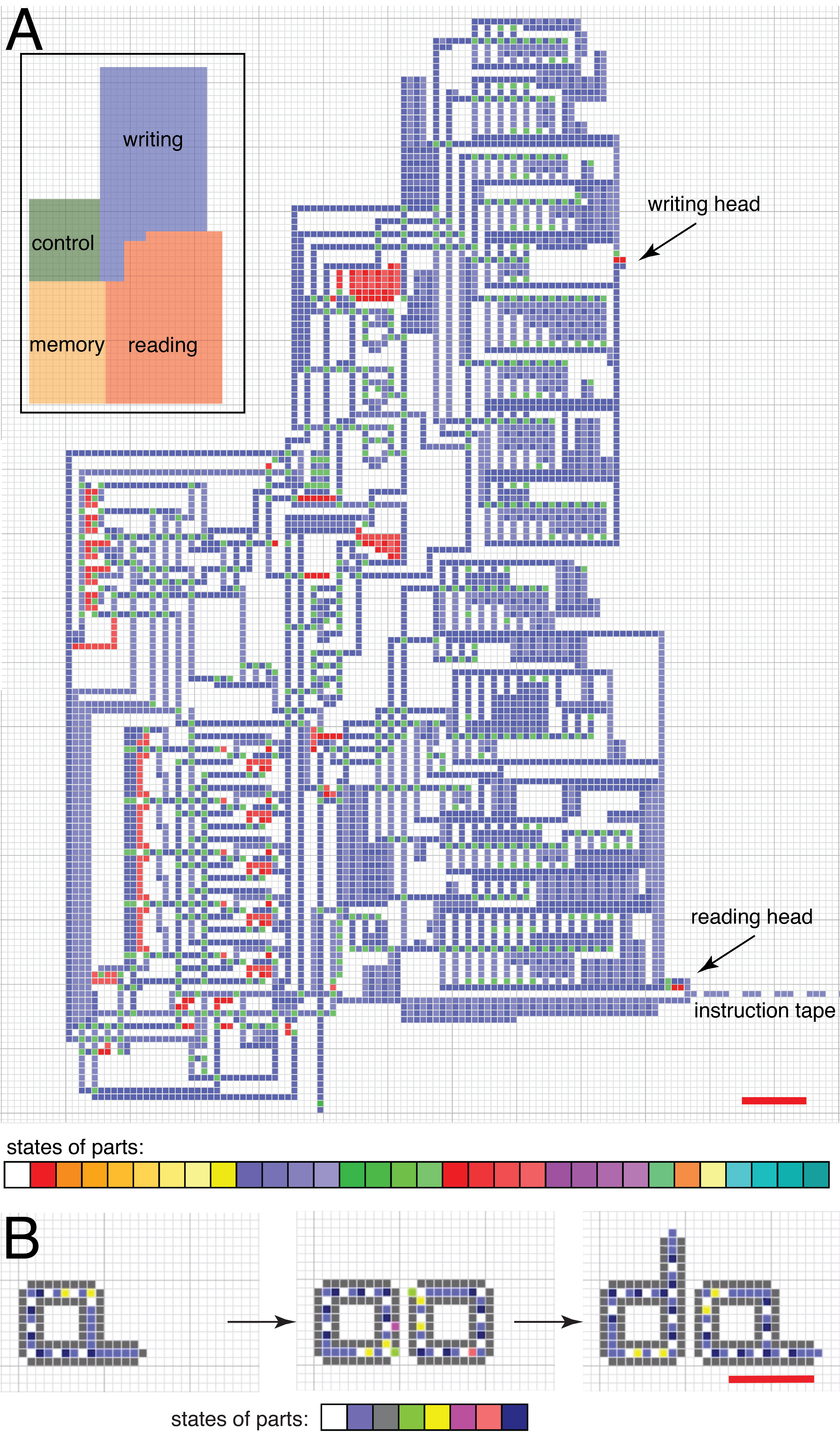}
  \caption{\small\textbf{Figure 1.} Self-replicating `machines' with instructions held in a static tape or in a dynamic tape have been implemented in cellular automata. (A) Implementation of John von Neumann's design of universal constructor \cite{Pesavento1995}. \textit{Top,} The universal constructor in the starting configuration. \textit{Inset,} Schematic of broad regions within the universal constructor. \textit{Bottom,} The 32 states of the parts that make up the machine (see \cite{Pesavento1995} for the meaning of each color). (B) Implementation of the Langton loop \cite{Langton1984}. \textit{Left,} The loop in the starting configuration. \textit{Middle,} A replication intermediate showing use of all states. \textit{Right,} Loops near the end of one round of replication. \textit{Bottom,} The 8 states used for replicating the loop (see \cite{Langton1984} for the meaning of each color). Red bar indicates scale for comparing (A) and (B). See methods for additional details.}
\label{fig:1}
\end{figure}

This scheme for making self-replicating machines avoids the infinite regress of instructions stored within instructions and is often presented as analogous to the process of self-replication in living systems with the `instructions' being held in DNA. However, the instructions for replicating a machine can be held either in static tapes (e.g., the von Neumann universal constructor, Figure 1A) or in dynamic tapes (e.g., the Langton loop, Figure 1B).

These two different types of instruction storage can be viewed as occuring simultaneously in living systems as the `static' genome sequence and the `dynamic' recurring arrangement of molecules \cite{Jose2018}.

The transmission of the genome sequence from one generation to the next occurs along a lineage of cells that each go through cell division cycles. As a result, numerous additional cycling stores can carry information across generations. For example, the information for copying a genome is stored in an arrangement of molecules that changes during replication such that the genome is usually copied with a period of one cell division cycle. Additional cycling stores of information are clearly recognizable in oscillations that occur at different temporal and spatial scales relative to generation time. These include oscillations in the chemical modifications of molecules (e.g., a $\sim$24 hr period in cyanobacteria \cite{Rust2011} despite a $\sim$12 hr cell cycle \cite{Mori1996}), in the localizations of molecules (e.g.,  a $\sim$40 s period in cell lines that have a $\sim$24 hr cell cycle \cite{Tsai2005}), in the collective morphology of embryonic cells (e.g., a $\sim$0.5 hr period in sea anemone that have a $\sim$0.5 hr cell cycle \cite{Ragkousi2017}), and in the activity levels of circuits (e.g., $\sim$24 hr circadian rhythms in non-cycling neurons \cite{Hastings2018}).

From these considerations, the following realizations emerge about living systems:

(1) The transmission of form and function across generations relies on many stores of information that cycle with different periods that could each in principle range from less than the duration of one cell division cycle to more than that of one generation.

(2) The relative phases of the many cycles within the continuous lineage of cells between generations creates distinct states over time such that the cell code for the development of an organism is approximated at the start of each generation.

Thus, the integrated process of self-replication cannot be artificially parsed into the static genome that holds all the instructions to be interpreted by the dynamic molecular machines in the cell.

\vspace{5pt}
\noindent\textbf{Information in self-replicating machines}
\vspace{5pt}

Consideration of the total information stored in a self-replicating machine can clarify the different stores of information required for replication and sharpen the corresponding unknowns in living systems. For example, consider the self-replicating universal constructor (Figure 1A), which has a `machine' that has 6,329 parts with 32 states per part and uses an instruction tape that has 145,315 parts with 2 states per part. The maximal information stored in this machine could be enumerated by separately considering three different stores that each have analogs in living systems: (1) the configuration or shape of the machine, (2) the instruction tape, and (3) the parts of the machine.

The information stored in the shape of the machine is incalculably large because we have to consider the universe of shapes from which the particular assembly of parts that make the machine was selected (see supplemental text and Supplemental Figure 1 for a proof). This information is akin to the information required for getting together the particular collection of molecules that constitutes each current living system and has accrued since before the origin of life along the lineage of every living system. Because the unknown information in all historical environments (i.e. past available complements of molecules) needs to be taken into account to determine what life accrued bit by bit \cite{Joyce2012}, the magnitude of this information is incalculable.

The maximal information that can be stored in the instruction tape that has $N = 145,315$ parts with two states each is given by $H = log_2 2^{N} = N = 145,315$ bits. This store is analogous to the linear genome where the information is stored in the sequence of the four bases in DNA (A, T, G, C). For such a genome of length $L$, $H = log_2 4^{L} = 2L$ bits.

The maximal information that can be stored in the machine that has $N = 6,329$ parts with 32 states each is given by $H = log_2 32^{N} = 5N = 31,645$ bits. This store could be analogous to everything other than the genome sequence within the bottleneck stage. However, unlike in cellular automata, discrete states of living systems are not easily defined. Calculating the information in this potentially vast store requires definition of the biologically relevant states of the bottleneck stage. For any given genome, knowing the rest of the cell code for different organisms is a prerequisite for constructing living systems of varying complexity.

Here, I develop a framework for all the heritable information in a living system in terms of what that system can sense about itself and its environment. This framework is useful for guiding the experimental analysis of living systems and potentially for the design and analysis of other persistent adapting systems.

\vspace{5pt}

\noindent\textbf{Heritable information in living systems}

\vspace{5pt}
The spatial arrangement of the genome and everything else within the bottleneck stage could change over the course of development such that similar arrangements are reached with a period of one life cycle. As a result, molecules that are part of the recurring cell code could play different roles throughout development and defy permenant classification based on their roles. For example, an abundant maternal RNA that is simply used as a source of nucleotides in the developing embryo could at a later stage become a message that is translated into a protein. Nevertheless, a temporary classification during the bottleneck stage is necessary to enumerate the bits of information stored in cell codes. To facilitate this enumeration in units that are relevant for each living system and its environment, I propose considering entities, their sensors, and the sensed properties.

\textit{Entities.} An entity is a molecule\footnote[3]{For simplicity, the term molecule is used to refer to everything found in a living system that is chemically isolatable such as ions, atoms, and chemically bonded collections of atoms, and is extensible to all factors that remain to be discovered.} or association of molecules within a living system or in the environment that interacts with the living system. A cell code can include entities that are measured through interaction with other entities sometime during the life cycle and also entities that are never measured, which can be considered as byproducts made by the processes of life. Such effectively inert and unmeasured entities could nevertheless non-specifically contribute to molecular crowding at the bottleneck stage and thereby affect interactions among other entities. While the number of all molecules in a cell is large but countable, the combinatorial associations of molecules could make the total number of effective entities ($N$) larger still. Cellular components that are entities or parts of entities include small molecules such as ATP, water, ions, metabolites, etc., for which perhaps only concentrations are discerned by sensors, and macromolecules such as individual proteins, RNAs, polysaccharides, etc., for which concentrations, sequences, and conformations may all be discerned by sensors.

\textit{Sensors.} A sensor is an entity or an association of entities within the living system that responds to changes in other entities with changes in its properties such that these changes can result in subsequent changes in the rest of the living system or its environment. A sensor could sense entities within the system ($N$ total) or in the outside environment ($O$ total) that interacts with the system (e.g., salts, nutrients, etc.). An entity that binds or collides with another entity without any specific downstream consequences is not considered a sensor (e.g. one water molecule bumping onto a membrane). An entity could be a part of multiple sensors. For example, a protein complex formed by the association of A, B, and C proteins could be detecting and responding to the concentration of ATP while another protein complex made of A and C could be detecting GTP. Conversely, multiple sensors could be measuring the same entity. For example, the many kinases in the cytosol are all potentially sensitive to the levels of a common pool of ATP. By these definitions, ATP itself can be a sensor because its levels change in response to production by a synthase and this change is communicated to the kinases that respond to changes in ATP levels. All sensors are entities, but not all entities are sensors.

\textit{Properties.} A property is an attribute of an entity that is relevant for a living system because a sensor exists that can respond to changes in the values of that attribute. The number of different values for a property of an entity depends on the sensor and on the regulatory constraints of the system. Consider two sensors that can detect changes in the number of molecules of a particular RNA: a protein Lo responds when the numbers increase by 10 and a protein Hi responds when the numbers increase by 100. These two proteins would thus each `see' different numbers of measurable units for the same property (number of molecules) of the same entity. However, not all detectable values for a relevant property of an entity could be attainable because of the regulatory constraints of the system. For example, if the RNA accumulated in steps of 50 molecules at a time, then many of the values measurable by Lo are never available in the living system because the system changes in steps that are larger than the measuring step of the Lo sensor.

\textit{Environment.} Organisms develop as open systems interacting with the environment. Therefore, some entities in the environment are measured and reacted to by the living system throughout development. Even for a constant environment, some entities may be measured by different sensors that are active at different times during development. As a result, living systems are really system-environment combinations. Aspects of `sensing' considered for interactions within the system are also relevant for interactions with the environment. Specifically, the sensed attributes of entities in the environment depend on the nature of the evolved sensors in the system and molecular crowders in the environment can modify the interaction between sensors in the system and entities in the environment.

\textit{Forces.} Living systems can generate and be exposed to many kinds of forces, which are not being explicitly considered in the framework for heritable information developed here. Rather, they are implicitly accounted for in the properties of entities. For example, an entity experiences/exerts gravity because of its mass, electrostatic attraction because of its charge,  tension because of its elasticity, etc. Thus, heritable information captured using entities, sensors, and properties can account for relevant forces in the living system or in the environment.

\textit{Configurations.} The number of ways in which molecules can be arranged in the bottleneck stage such that they can be distinguished by the system provides an upper bound for the information that can be stored in a cell code, which is the subset of configurations that are nearly reproduced during the bottleneck stage of successive generations. The maximal number of such distinguishable configurations of a living system for a given number of interacting entities in the environment is given by the product of the number of possible genome sequences and the number of possible systems and their coupled environments that can support each genome sequence. Assuming that each system-environment combination generates one characteristic set of unmeasured entities that contribute to crowding effects, the number of distinguisable configurations for a living system and its environment during the bottleneck stage ($C_{tot}$) is given by:
\begin{align*}
C_{tot} &= \text{genomes } \times \text{ system-environments}\\
C_{tot} &= X^L(\sum_{i=1}^{B}E_i(\sum_{j=1}^{S_i}S_{j}(\sum_{k=1}^{P_{j}}P_{k})))\hspace{30pt}(1)\\
X &= \text{number of types of bases in the genome.}\\
L &= \text{length of the genome in base pairs.}
\end{align*}
\begin{align*}
	E &= \text{measured entity (total }B\text{ in the bottleneck}\\
	&\text{\hspace{15pt} stage: }N_b\text{ in system, }O_b\text{ in environment).}\\
	S &= \text{measuring sensor (total = }S_i \text{ for }i^{th} \text{ entity).}\\
	&= f(Y)\text{, where each } Y\subseteq\{E_1, E_2, ...E_N\}, \text{i.e. a}\\
	&\text{\hspace{15pt}configuration of entities, }N\text{ per life cycle.}\\
	P &= \text{attainable and measurable values of property}\\
	&\text{\hspace{15pt}(total = }P_{j}\text{ for }j^{th}\text{ sensor of } i^{th} \text{ entity).}
\end{align*}

This Entity-Sensor-Property framework enumerates all distinguishable configurations as a product of four terms that encapsulate the maximal numbers of distinct states in two stores of information: $X^L$ enumerates all possible genome sequences, which are replicating stores of heritable information, and $\sum_{i}E_{i}\sum_{j}S_{j}\sum_{k}P_{k}$ enumerates all potentially recurring arrangements of interacting molecules, which are cycling stores of heritable information. Such enumeration without considering rearrangements of chemical bonds within any molecule can be thought of as biological entropy and is less than the chemical entropy of an organism, which was initially estimated allowing for rearrangements of chemical bonds to be $\approx4.2\times10^{10}$ bits for \textit{E. coli} \cite{Morowitz1955}.

It is clear that the replicating store cannot uniquely predict the cycling store as evidenced by most distinguishable cell types of the human body all having the same genome sequence. However, interdependence of the two stores and compatibility with the perpetuation of life reduces this maximal number of distinguishable states of the bottleneck stage. In other words, fewer configurations can act as heritable cell codes ($Cell\hspace{3pt}Code_{tot}<C_{tot}$) because of mutual constraints between the arrangement of molecules and the genome sequence in living systems. First, some genome sequences may not be sufficiently complex to support any living system (e.g. a genome of all As, all Gs, all Cs, or all Ts.). Second, each genome sequence \textit{constrains} but does not dictate the number and kinds of entities that could be part of any cell that contains the genome (e.g. DNA sequence constrains RNA sequence, which constrains protein sequence). Third, the genome sequence may also constrain the total number of possible arrangements of molecules within any cell - i.e., the number of cell states and cell types - in a given environment. Fourth, the lineage of cells that connects two generations may be incapable of supporting some cell types because of the need to return to the cell code at the start of each generation within the context of a living system (i.e., some differentiated cell types may be irreversible within the context of the living system although many can be transformed into pluripotent stem cells in vitro \cite{Yamanaka2012}). The number of all possible cell codes, however, is likely greater than that seen in evolved organisms because the historical process of evolution is not expected to allow exploration of every cell code (i.e., $C_{tot}>Cell\hspace{3pt}Code_{tot}>Cell\hspace{3pt}Code_{evol}$).

Cell codes of varying complexity have evolved over time to specify the development of each organism that has ever existed \cite{Jose2018}. Cell codes could in principle differ in the relative amounts of information stored in the genome sequence versus in the arrangement of molecules. The interdependence of these two stores of information invite exploration of the relationship between their storage capacities over evolutionary time. Consider the consequences of adding into a pre-existing cell code a newly evolved gene sequence that codes for a protein. The number of possible genome sequences of a given length that can support this cell code decreases because fewer distinct genomes can include the gene sequence for the new protein (i.e., total sequences becomes less than $X^L$, Supplemental Figure 2). However, the number of distinguishable arrangements of all molecules can either increase or decrease. Increase can occur because addition of the new DNA sequence, the transcribed RNA, and the translated protein to the contents of the cell could all lead to new interactions with pre-existing molecules (i.e. $E$, $S$, and $P$ could all increase, resulting in a larger value for $\sum_{i}E_{i}\sum_{j}S_{j}\sum_{k}P_{k}$). Decrease can occur because these new molecules could constrain the arrangement through regulatory interactions (see section titled Entity-Sensor-Property: insights). Furthermore, the magnitude of changes in $\sum_{i}E_{i}\sum_{j}S_{j}\sum_{k}P_{k}$ depends on the nature of the new gene product (e.g., expression or repression of many gene sequences by a transcriptional activator or repressor, respectively, could lead to large changes). Studies on the origin and evolution of information storage could illuminate trends in the partition of heritable information between different molecular stores and lead to general principles (for related views emphasizing arrangement see \cite{Walker2013, Fields2018}, genome sequence see \cite{Lynch2007, Szathmary1995}, anatomy see \cite{Levin2020}, and energy see \cite{Annila2014}). For example, the complexity of cell codes, and thus organisms, may have increased through restriction of the genome sequence along with expansion of the arrangement of molecules as sources of neutral or adaptive variation.

\vspace{5pt}

\noindent\textbf{Entity-Sensor-Property: extensions}

\vspace{5pt}
Several processes in living systems could limit or expand the number of arrangements in the bottleneck stage ($\sum_{i}E_{i}\sum_{j}S_{j}\sum_{k}P_{k}$). Processes that can change the information content of cell codes by decreasing (e.g., self-organization and self-assembly), increasing (e.g., chemical modification) or variably changing (e.g., compartmentalization) entities, sensors, and/or properties are being actively analyzed. Living systems could manipulate heritable information through the regulation of all such processes.

\textit{Impact of self-assembly and self-organization. }Order can arise through the spontaneous association of molecules in living systems. Two forms of such spontaneous order have been recognized: (1) self-assembly, which refers to the formation of static structures that are relatively stable (e.g., viruses, flagella)\cite{Kushner1969}; and (2) self-organization, which refers to the formation of dynamic structures that appear stable (e.g., cytoskeleton, endocytic compartments) \cite{Nicolis1977}. Both forms of order, however, depend on the immediate molecular environment. Therefore, changing the surroundings of a `self-assembled' or `self-organized' structure can result in alternative configurations that may be distinguishable by evolved sensors. For example, cells can use an adaptor protein to modulate the size of vesicles that form through self-assembly \cite{Zhang1998} and cells can respond to pressure by reversibly disassembling the mitotic spindle that is maintained through self-organization \cite{Salmon1975a, Salmon1975b}. In this way, living systems can store and retrieve information from self-assembled and self-organized collections of molecules.

\textit{Impact of chemical modifications. }Modifications of nucleic acids ($^{5m}$C, $^{5hm}$C, $^{m6}$A, etc.) or proteins (phosphorylation, methylation, glycosylation, etc.) result in new entities with properties that could potentially be measured by sensors. Modified bases on the genome could increase the number of possible spatial arrangements of the genome and its binding partners (i.e., $E$, $S$, and $P$ in equation(1)), and could also increase sequence information (i.e., $X$ in equation (1)) \textit{if} the modification alters base-pairing. Modifications on RNA or proteins on the other hand could either increase or decrease $E$, $S$, and $P$, but always reduce the maximal number of genomes of a given length that could support such a modification because each possible genome would be constrained to include the gene sequence for the enzyme that catalyzes the modification (i.e. total sequences become less than $X^L$). Similar considerations hold for modifications of all other molecules in the bottleneck stage.

\textit{Impact of compartmentalization. }Living systems dynamically manipulate which entities come together into organized units and which outputs from these units are subsequently measured. When different subcellular compartments form, the same entity or sensor could be present in two or more different compartments. If two such pools of the same entity are sensed separately during the life cycle of an organism, the total number of possible configurations are effectively increased. Alternatively, many different entities could be encapsulated into one compartment. If only a few aggregate properties of the compartment are sensed during the life cycle of an organism (e.g. droplet sizes of phase-separated aggregates such as RNA granules \cite{Shin2017} or numbers of organelles such as mitochondria), the number of distinguishable configurations are effectively reduced.

These different ways of changing entites, sensors, and properties highlight the multiscale nature of living systems and suggest the utility of different Entity-Sensor-Property frameworks at different scales and across scales.

\vspace{5pt}

\noindent\textbf{Entity-Sensor-Property: insights}

\vspace{5pt}

To appreciate some implications of the framework, consider a toy model where the genome sequence and the environment are held constant (Figure 2).

Let the remaining contents of a `cell' include three entities ($E_1, E_2, E_3$ - three english letters) that can be at four different states (two fonts with upper and lower cases) and be sensed by two sensors ($S_1$ measuring lines and $S_2$ measuring curves). Each state is analogous to different \textit{experimentally} measurable values for a property of molecules in a cell (e.g., concentration, localization, shape, charge, etc.). Consider the entity $E_1$ in state `{\fontfamily{phv}{\selectfont A}}' made of three straight lines. A sensor that measures lines could measure one of numerous possible properties: thickness of lines, color of lines, length of lines, etc. For simplicity, let number be the only property $P$ sensed by both $S_1$ and $S_2$. For example, the value of the property sensed by $S_1$ of $E_1$ in state `{\fontfamily{phv}{\selectfont A}}' is 3 and that sensed by $S_2$ of $E_3$ in state `{\fontfamily{phv}{\selectfont c}}' is 1 (see Figure 2 for all values).

\textit{Sensors can limit information storage. }To calculate the relevant information stored in a system, we need to enumerate the number of different states of the entities sensed by the system (Figure 2).
\begin{figure}[H]
  \centering
  \includegraphics[width=\linewidth]{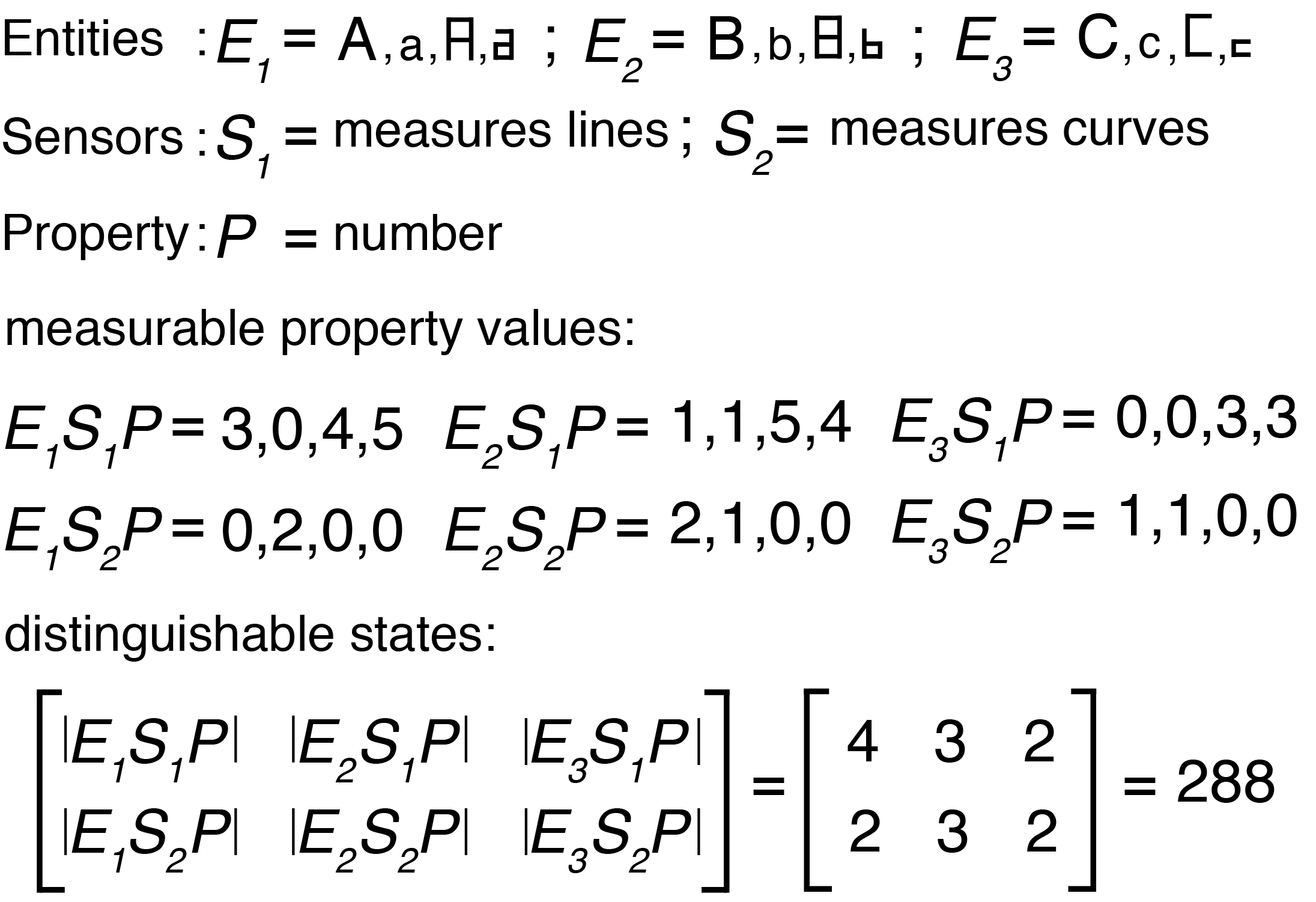}
\caption{\small\textbf{Figure 2.} Distinguishable states in a toy model of possible cell codes with a given `genome' and `environment'. Three entities ($E_1, E_2, E_3$), two sensors ($S_1, S_2$), and one sensed property ($P$) are considered. The measurable property values of each entity by each sensor is enumerated ($E_1S_1P, E_2S_2P, ...$). Each distinguishable set of property values for all entities defines a distinguishable state. Therefore, the number of distinct elements in a set of the measured values (i.e $\left|E_1S_1P\right|, \left|E_1S_2P\right|, ...$) can be used to calculate the total number of distinguishable states in the system ($4\times3\times2\times2\times3\times2 = 288$), which is less than the number expected if every value of every entity were distinguishable ($4\times4\times4\times4\times4\times4 = 4096$).}
\label{fig:2}
\end{figure}
While each sensor can sense one property of each entity, different states of an entity may not always be distinguishable by a sensor (e.g. $S_1$ will measure the states `{\fontfamily{phv}{\selectfont C}}' and `{\fontfamily{phv}{\selectfont c}}' of $E_3$ as 0 and $S_2$ will measure both as 1). Such indistinguishability is evident in living systems as the requirement for threshold levels of a signal for a detectable response. Thus, `thresholding' by the sensor results in a reduction in the total number of states of the system and thus the storable information (Figure 2). For example, this system of $E$, $S$, and $P$ can only distinguish 288 states ($\approx$8.17 bits), but a naive estimate based on the ability to distinguish all states of all entities yields 4096 states (12 bits). Thus, the number of entropic states of a system depends on the available sensors and their sensitivity.

\textit{Distinct states may be equivalent. }Selection can impose external constraints on the form and function of a living system. For example, the environment of the system in the toy model might require `cells' with consistent fonts and case for survival. This would result in the survival of cells with `{\fontfamily{phv}{\selectfont ABC}}', `{\fontfamily{phv}{\selectfont abc}}', etc. as the values for each of the three entities (Supplemental Figure 3). However because there are two possible cases (upper vs. lower), there would be two distinguishable states that are effectively equivalent for survival in this environment. Such situations could result in unregulated redundancy such that similar functions are performed by different molecules in random sets of cells \cite{Ravikumar2019}. Over evolutionary time scales, this type of unregulated redundancy could result in organisms with similar form and function but different underlying molecular mechanisms \cite{True2001}. These considerations also hold when such equivalency is imposed by sensors that fail to distinguish different entities. For example, a channel protein responding to changes in membrane potential would measure changes in different ions as equivalent as long as the end result was a similar change in potential \cite{Levin2012}.
\begin{figure}[H]
  \centering
  \includegraphics[width=\linewidth]{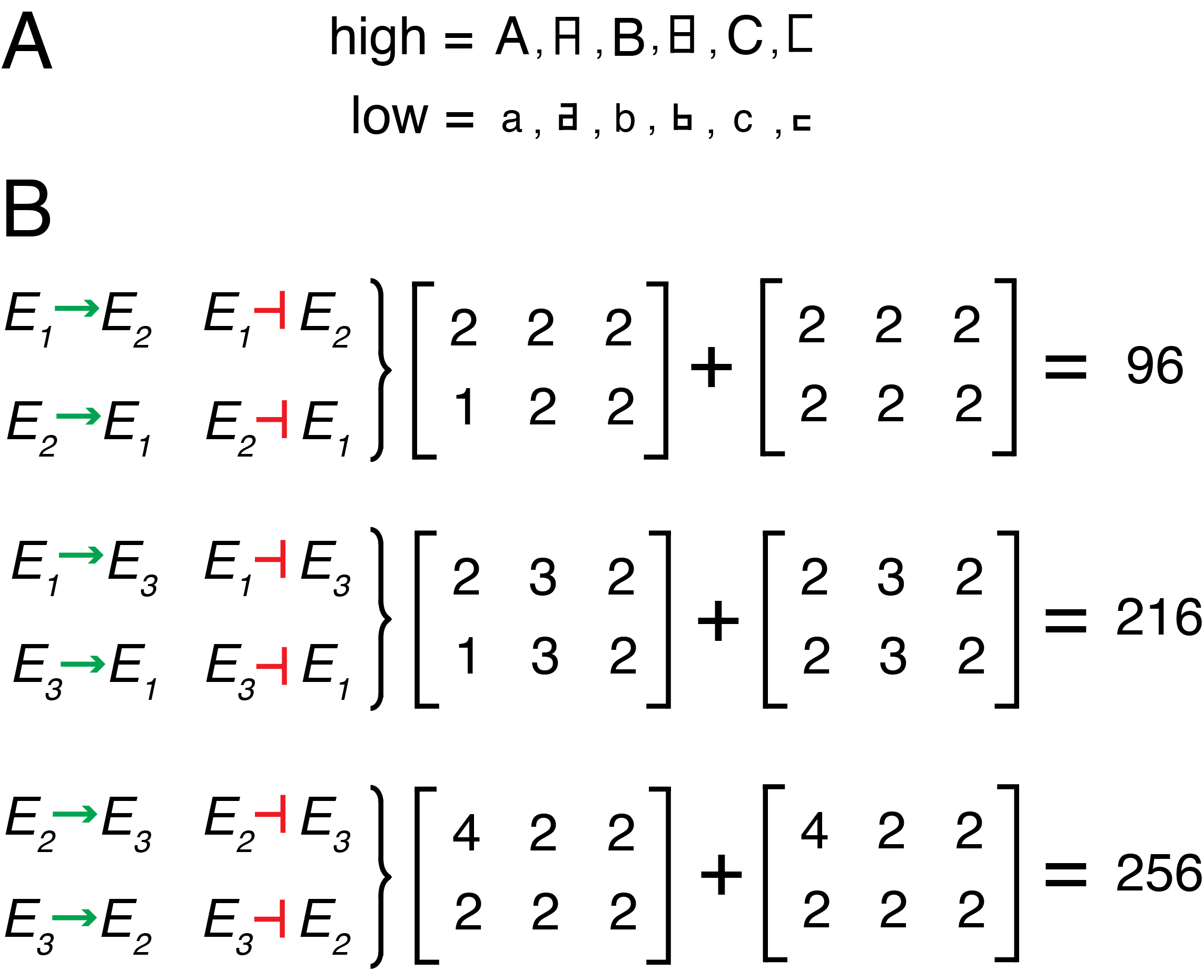}
  \caption{\small\textbf{Figure 3.} Regulation reduces the number of states that can be sensed by the system. (A) Two states for each entity - high (upper case) or low (lower case) - were considered for exploring the impact of regulation in the toy model. (B) Consequences of introducing regulatory constraints. Inhibition (bar) or activation (arrow) between all pairs of entities were considered. Matrices of distinguishable values (product of all = number of states) for each cell with regulatory interactions between $E_1$ and $E_2$ (\textit{top}), $E_1$ and $E_3$ (\textit{middle}), or $E_2$ and $E_3$ (\textit{bottom}) are shown. Different regulatory constraints result in differential reduction in the number of states of the system.}
\label{fig:3}
\end{figure}
\textit{Regulation reduces sensed states.} The different states of each entity could be classified as high (upper case) or low (lower case) to simplify the analysis of regulation in the system (Figure 3A). This simplification is similar to Boolean networks that have been used to explore the impact of regulation \cite{Kauffman1969}. All additions of either activation or inhibition as a regulatory interaction between two entities reduces the number of distinguishable states in the system (Figure 3B). This reduction occurs because any regulatory interaction between two entities couples changes in those entities. As a result, two entities that were previously free to vary independently become either directly or inversely correlated, leading to an overall reduction in the number of possible states. Different regulatory architectures can lead to different states with equivalent capacity for information storage. Specifically, 12 different single regulatory interactions in the toy model lead to only 3 different storage capacities - 96, 216, or 256 states (Figure 3B). Adding two regulatory interactions results in all 36 different regulatory architectures having only 96 distinguishable states (Supplemental Figure 4). These results suggest a preliminary conclusion: regulation reduces the ability of systems to store information in the arrangement of molecules.

\textit{Reducing states may promote robustness. } Robustness is the ability of living systems to remain similar despite \textit{some} variation introduced by environmental or internal conditions \cite{Stelling2004}. In other words, some changes either do not alter anything about a robust system or can alter some entities but nevertheless do not substantially affect the system. The differences in the number of states in cells with different regulatory architectures (Figure 3B) suggest a relationship between regulation and robustness of cell types. Unlike in the toy model, in living systems, all sensors are made from entities (equation (1)). Therefore, cell types with fewer states could be more robust because they are only capable of sensing, and thus responding to, fewer perturbations. Conversely, cell types with many states could be less robust because they are capable of sensing and responding to many perturbations. Changes in regulatory architectures could therefore be used to generate cell types that are differently responsive to external signals, which may have implications for the observed robustness of development \cite{Waddington1942}. To achieve such robust development, entities need to be assembled into cell codes such that the same sequence of events unfolds despite some perturbations. Storing entities as perturbation-resistant assemblies or combining entities that fail under some conditions with entities that fail under other conditions (redundancy) are possible ways to ensure robust cell codes and subsequent development. An additional possibility suggested by these observations is reducing the number of sensors through increased regulation such that some perturbations are simply not sensed.
\vspace{5pt}

\noindent\textbf{Two-base genomes could be part of efficient living systems}

\vspace{5pt}
Our current ability to exquisitely edit genomes and transcriptomes \cite{Knott2018, Rees2018} is a limited manipulation of living systems in that the outcome of the edit is entirely determined by how the living system interprets the change. In other words, we can make changes to a sequence and read out what the living system does with the changed sequence but we cannot yet make changes that instruct a living system to perform arbitrary tasks. Such expanded manipulation could require ways of increasing the complexity of the stored heritable information. As suggested by equation (1), this increase could be achieved by either increasing storage in the genome sequence or by increasing storage in the arrangement of molecules. Increases in the storage capacity of a genome by increasing the number of different bases will require concomitent increases in the complexity of the machinery for accurate reading and writing of the genome. For example, a 16-base genome of length $L$ has four times the capacity of a 2-base genome ($4L$ bits vs $L$ bits).
\begin{figure}[H]
  \centering
  \includegraphics[width=\linewidth]{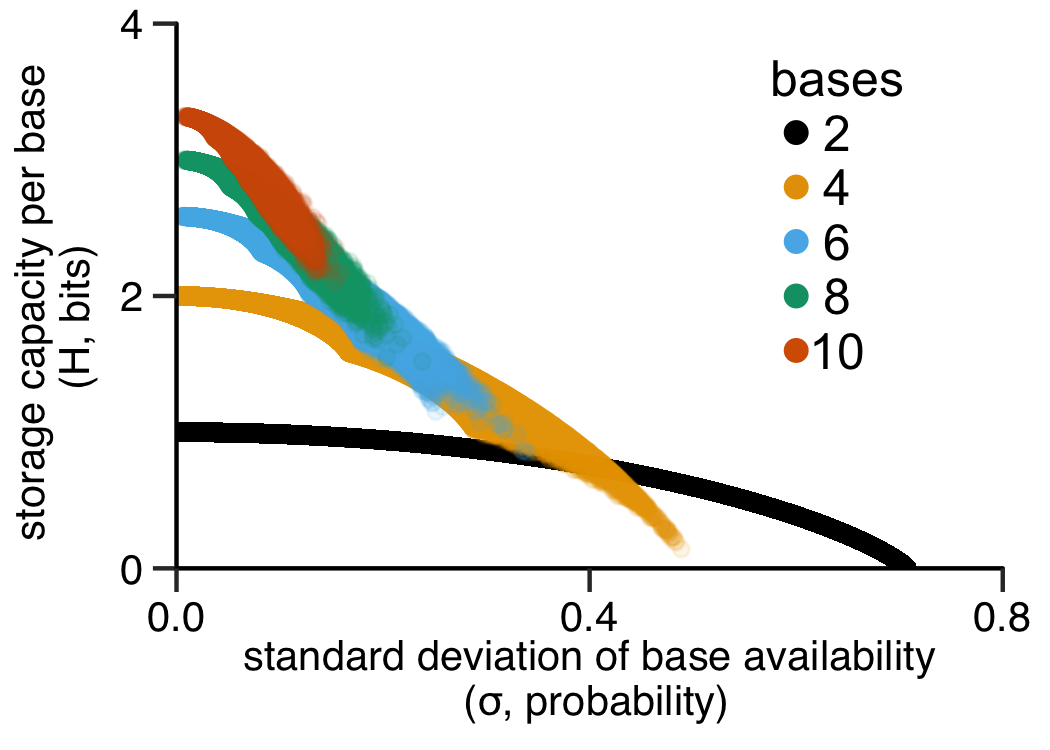}
  \caption{\small\textbf{Figure 4.} Increases in the types of base pairs in a genome increase the maximal information that can be stored per base but decrease the mean variation in base availability that can support the high information storage. Plot showing how maximal information ($H$) of a base pair varies with the standard deviation ($\sigma$) of base probability. For each n-base system (n = 2, 4, 6, 8, 10), the results for one million base probabilities drawn form a uniform distribution are plotted.}
\label{fig:4}
\end{figure}
However, such a genome would require machinery for discerning eight times as many kinds of bases. Furthermore, the range of availabilities of bases that can support the enhanced information storage decreases with an increase in the number of different bases (Figure 4).

Perhaps the simplest route to the synthesis of living systems with arbitrary capacity to store heritable information would be to use a longer 2-base genome that is equivalent to the natural 4-base genome (e.g. needs 5-base codons for encoding at least 20 amino acids, $2^5 > 20 > 2^4$), but can be supported by simpler machinery in the cell to read and write the genome. Testing the practicability of this speculation requires systematically changing the chemistry of the genome \textit{and} the cell while preserving overall storage capacity.

\vspace{5pt}
\noindent\textbf{Discussion}
\vspace{5pt}

By jointly considering all information transmitted from one generation to the next using molecules, I have developed an expanded view of heredity (see supplemental text for other applications). Heritable information stored outside the genome sequence is limited by mutual constraints with the sequence, by regulatory architectures, and by what a living system can sense about itself and its environment.

\textit{Strengths and limitations of framework.} Entities and sensors in a cell were parsed based on their roles at a particular time in development - the bottleneck stage. However, the roles of entities and sensors are potentially interconvertible over time. A sensor could become a unresponsive entity for a while and an entity could become a responsive sensor when it encounters another appropriate entity. Such changes in roles are likely part of the changes during development that lead to the assembly of cell codes at the start of each generation. Given this time-bound nature of entities and sensors, what is the \textit{duration} of a bottleneck stage? This question is currently very difficult to answer and poses a practical problem for unambiguously defining the cell code of an organism. Nevertheless, the stability of cell types suggests that functionally important states are preserved for significant periods through homeostasis.

The framework presented here does not account for the stochastic and noisy nature of all interactions within a cell. For example, there are fundamental limits to control that result from information loss \cite{Lestas2010} and the physical limits of biochemical signaling \cite{Berg1977, Bialek2005}. Unlike in man-made communication systems, the presence of numerous simultaneous signaling pathways in living systems - including as yet unknown pathways - makes it unclear whether any observed variation in one signaling pathway should be characterized a priori as interference from another signaling pathway or as noise. Nevertheless, developing an understanding of heredity in terms of genome sequence, entities, sensors and properties is a first step towards future extensions of the framework that could address these issues.

Some past frameworks for analyzing living systems provide conceptual structures for explaining their evolution and behavior but do not inform their construction or origin. Models that analyze evolutionary outcomes regardless of the material basis of genotype and phenotype (e.g., ref. \cite{Rivoire2014}) are useful guides for the analysis of organisms at the population level but not for the construction of organisms from molecules sought here. Phylogeny, architecture, and adaptation have been combined to understand trends in the evolution of form \cite{Briggs2005}, but such models are currently not fine-grained enough to enable construction. The productive analysis of complex systems by partitioning a system into abstract nodes and edges to view particular aspects of living systems as networks \cite{Barabasi2004} has generated intuitions and approaches that could be extended to the framework presented here. Such extension beyond abstract networks is necessary to enable the construction of living systems because typical abstractions do not incorporate all relevant properties of cellular contents. The explicit consideration of relevant properties for all entities that are measured by sensors in the framework presented here could help in accruing knowledge in a form that is useful for the construction of living systems and for the realization of a practical systems biology \cite{Kitano2002a, Kitano2002b}.

\textit{Synthesis of living systems.} Building something using its constituent parts is a good way to discover the flaws in our understanding of how it is put together. For example, it is currently unclear if \textit{perfect} self-replication ever occurs in living systems. Perhaps the perpetuation of life is always associated with having entities that are not recreated with a period of one generation but rather with longer or shorter periods. For example, when the noisy and variable behavior of a synthetic oscillating circuit in \textit{E. coli} \cite{Elowitz2000} was improved to obtain synchronous long-term oscillations \cite{Potvin2016}, the period of oscillation increased to 14 generations. Such possibilities can be explored by allowing different generation times for the precise recreation of some entities and arrangements in the cell code. The similarity in form and function of parent and progeny, however, suggests that the cell codes recreated with a period of one generation are at least nearly equivalent for specifying development in each generation.

Evolved cell codes are unlikely to be efficient stores of heritable information because of the historical measures and counter-measures through which evolution proceeds \cite{Jacob1977, Gould1979}. Efficient storage of the mutual information between two variables can be achieved using a compressed bottleneck variable \cite{Tishby2000, Kolchinsky2018}. If there was selection for effectively packing maximal information into the bottleneck stage in living systems, the entities and arrangements of evolved cell codes could similarly be efficient stores of the mutual information between the past and the future. All such efficient cell codes might have similar characterstics as observed in cellular automata in which the capacity to support computation emerges (captured in the $\lambda$ parameter in \cite{Langton1990}). Despite the possibility of such overall optimization, it is unclear if living systems can evolve to maximally optimize information storage and/or transmission for a particular trait. In fact, it might be difficult to define what the `optimum' is for a process because the presence of many homeostatic mechanisms in cells, including transgenerational homeostasis \cite{Jose2018}, require opposing processes that could limit optimality. Experimental approaches that attempt to generate minimal bacterial cells \cite{Hutchison2016} need to be extended to different organisms to discover how the complexity of organisms scales with their cell codes.

Making efficient living systems of arbitrary complexity requires a holistic approach to information handling. The joint cosideration of all heritable information presented in this article suggests that a genome with two different kinds of bases might function as an efficient replicating store when combined with the simplest possible cycling stores (Figure 4). Thus far, experimental approaches to fundamentally change heritable information have focused on increasing the storage capacity of the genome. A 50\% increase in the storage capacity of DNA sequence can be achieved by doubling the number of different bases in DNA \cite{Hoshika2019}. Furthermore, an organism that uses a 4-base genome can be modified with two additional DNA bases to successfully store \cite{Zhang2017a} and retrieve \cite{Zhang2017b} information. In contrast, we cannot yet engineer such increases in the information stored by the arrangement of molecules because our knowledge of this store of heritable information is in its infancy. The theoretical and practical limits of varying \textit{all} heritable information deserve exploration to understand the evolution of natural, modified, and synthetic living systems.
\vspace{5pt}

\noindent\textbf{Acknowledgements}

\vspace{5pt}
I thank Tom Kocher, Karen Carleton, Charles Delwiche, David Wolpert, Chris Kempes, Michael Lachmann, Artemy Kolchinsky, James Yorke, Daniel Damineli, Pierre-Emanuel Jabin, Katerina Ragkousi, David Jordan, KP Mohanan, LS Sashidhara, Sudha Rajamani, Jyotsna Dhavan, Michael Levin, Alejandro S\'anchez Alvarado, Ajay Chitnis, and Victor Ambros for discussions and encouragement; and Tom Kocher, Karen Carleton, Ken Helland, members of the Jose lab, and an anonymous reviewer for comments on the manuscript. Research in my lab is supported by the NIH (R01GM111457 and R01GM124356).
\vspace{5pt}

\noindent\textbf{References}

\vspace{5pt}
\printbibliography[heading=none]
\end{multicols}
\end{document}


\begin{center}
\LARGE{A framework for parsing heritable information}\\
\vspace{10pt}

\large{Antony M Jose}\\
\small{\textit{Department of Cell Biology and Molecular Genetics,}}\\
\small{\textit{University of Maryland,}}\\
\small{\textit{College Park, MD 20742, USA}}\\
\vspace{100pt}

\Large\textbf{Supplemental Material}\\
\vspace{5pt}
\small{Text, 4 Figures and Figure Legends, Methods, and References.}
\end{center}
\pagebreak

\noindent{\Large{\textbf{Supplemental Text}}}

\vspace{5pt}

\noindent\textbf{Variations on the single-cell bottleneck}

\vspace{5pt}
The life cycles of many multicellular organisms include a single-cell stage that limits the transmission of heritable information. However, the amount of information transmitted through this bottleneck stage could vary based on the ecology and developmental strategy of particular organisms.

Growth in a predictable environment that is stable for many generations may reduce the information that needs to be transmitted through the bottleneck stage. The reliable association of microbiomes and other symbionts in each generation could similarly facilitate the reduction of information transmitted through the bottleneck stage. At an extreme, viruses and parasites effectively form joint systems of heredity and development with the organisms they infect.

Developmental strategies can impact the temporal and spatial reach of the information that is transmitted across generations. For example, in a female human fetus, the germ cell precursors that will generate the oocytes have already differentiated, which could facilitate communication from the pregnant mother to the unborn grandchild through shared circulation. Such expansions of the bottleneck stage may increase the complexity of heritable form and function.

While the framework for all heritable information developed in this study is presented for the common case of a single-cell bottleneck between generations, it applies for all alternative configurations of the bottleneck stage.
\vspace{5pt}

\noindent\textbf{Cell codes are assembled during development}

\vspace{5pt}

The form and function of organisms are mostly preserved from one generation to the next. This preservation requires that development begin in successive generations with similar genome sequences and similar arrangements of regulatory molecules. While the genome is copied during every cell division, the arrangement of molecules presumably changes during development but returns to a similar configuration after one life cycle. These recurring arrangements therefore are cycling stores of heritable information that along with the genome sequence form cell codes for the development of organisms in each generation \cite{Jose2018}.

All cells accumulate molecules using building blocks from the environment in ways that depend on pre-existing molecules within cells. This dependence on prior state means that the current `phenotype' of a cell is determined by the `genotype' \textit{and} the pre-existing phenotype of the cell. The observed similarity between organisms of successive generations implies that the bottleneck stage needs to have molecules arranged in such a way that similar temporal sequences and spatial patterns ensue during development in successive generations. In other words, cell codes encode both temporal and spatial order in spatial arrangement during the bottleneck stage.

Different collections and configurations of entities within a cell can arise through differences in physical and chemical processes that include: (1) the temporal sequence of binding or chemical reactions; (2) the relative rates of different reactions; (3) the confinement of reactions; (4) the amplification of biases that arise from intrinsic noise; (5) the addition of external entities; and (6) the destruction or secretion of entities. The arrangement of entities in a cell at any moment is an integrated consequence of the historical values of such differences leading up to that moment. An organism can therefore assemble the information for making a similar organism in the next generation by controlling such processes throughout development such that its bottleneck stage contains a well-configured cell code - a spatial representation of the past ready to shape the future.

\vspace{5pt}

\noindent\textbf{Information content of a shape}

\vspace{5pt}

To attempt calculating the information in a particular shape, we need to make assumptions about the universe of shapes from which that shape is drawn. For example, in a two-dimensional cellular automata environment, we could assume that the sets of parts from which the shape is formed are contiguous (i.e., each part shares at least one side with another part) and that rotations of shapes by multiples of 90 degrees are allowed. Given these assumptions, let ${S_1,.., S_n}$ be sets of parts that can each form one and only one target shape within it. The total number of objects of all sizes and shapes that could be formed using one such set of $S_i$ parts is given by $2^{S_{i}}$. Let  $O_{i}$ be each such set of $2^{S_{i}}$ objects (i.e. $\left|O_{i}\right| = 2^{S_{i}}$). The number of all uniquely shaped objects in all such sets combined is given by $\left|O_1 \cup O_2 ... O_{n-1} \cup O_n\right| = U$ - the maximal number of unique objects aggregated from universes that could each contain an object with the target shape once and only once. The maximal amount of information in the target shape is therefore given by $H = log_2{U}$ bits. Three simple cases illustrate how this number scales with the complexity and size of the shape in the cellular automata environment (Supplemental Figure 1).

For a target shape made of 1 part, $U = 2$ and $H = log_2(2) = 1 $ bit. For a target shape made of 2 parts that are next to each other, $U = 4$ and $H = log_2(4) = 2$ bits. However, for a target shape made of 3 parts that are next to each other in a row (or column), $U = \infty$ because an infinite number of parts that zig zag such that they only share a side with one row part and one column part could belong to the universe from which the shapes are drawn. Similarly, for all machines made of $n > 3$ parts in a row, contiguous sets of parts that zig zag and switch direction after fewer than $n$ parts can be constructed to make $U = \infty$. Thus, for the universal constructor made of 6,329 parts that are arranged into a complex shape (Figure 1A), the information content is incalculably large.

\vspace{5pt}

\noindent\textbf{Configurations of a gene sequence and its regulators}

\vspace{5pt}
Basic principles underlying the assembly of entire cell codes may be discoverable through reductionist studies on a few units of heredity. These studies would aim to discover how a unit of heredity is configured in one bottleneck stage, how that configuration changes during development, and how the starting configuration is recovered by the next bottleneck stage. More than a century of experimental analyses support the usefulness of analyzing units of heredity \cite{Jose2018}, which were initially called cell elements \cite{Mendel1866} and can be thought of as having two parts: (1) a gene sequence transmitted between generations as part of the genome sequence; and (2) gene regulators transmitted between generations as part of an arrangement of molecules. The precise limits of a gene sequence have proven to be difficult to establish \cite{Gerstein2007, Portin2017} and the precise configuration of \textit{all} regulators for a given gene is likely to be similarly difficult to establish. Nevertheless, formulating a unit of heredity that is associated with a gene sequence as a cell element is useful as a practical framework for reductionist studies. Such studies do not need to analyze all entities and their sensors - indeed this is impractical. Instead, subprocesses like transcription, translation, protein localization etc. could be partitioned and then the rest of the entities, their sensors, and sensed properties under study could be analyzed to determine if they are sufficient to account for the observed heritable phenomena. To facilitate such reductionist analyses, a gene sequence and its regulators could be parsed into a provisional entity-sensor-property system and other distantly interacting entities could be considered as part of the `environment'. After such simplification, the configurations that this provisional entity-sensor-property system can distinguish is given by:
\begin{align*}
c_{tot} &= x^l(\sum_{i=1}^{b}e_i(\sum_{j=1}^{s_i}s_{j}(\sum_{k=1}^{p_{j}}p_{k})))
\end{align*}
\begin{align*}
	x &= \text{number of different types of bases in the gene.}\\
	l &= \text{length of the gene sequence.}\\
	e &= \text{measured entity (total }b\text{ in the bottleneck stage: }n_b\text{ in system, }o_b\text{ in environment).}\\
	s &= \text{measuring sensor impacting regulation of the gene sequence (total} = s_i \text{ for }i^{th}\text{ entity).}\\
	&= f(y)\text{, where each } y\subseteq\{e_1, e_2, ...e_n\}, \text{i.e., a configuration of entities, }n\text{ per life cycle.}\\
	p &= \text{attainable and measurable values of property (total = }p_{j}\text{ for }j^{th}\text{ sensor of } i^{th} \text{ entity).}
\end{align*}

Cell elements that encode the expression patterns of gene sequences would therefore be subsets of ${c_{tot}}$ that are recreated in successive generations. Progressive application of this framework by considering larger systems successively could provide a principled approach for combining cell elements into cell codes.

\vspace{5pt}

\noindent\textbf{Other applications of the Entity-Sensor-Property framework}

\vspace{5pt}

The three parameters - entities, sensors, and properties - introduced here for measuring cycling stores of information may be applicable to broader non-biological classes of heritable information. As an extreme example, the persistence or evolution of ideas among groups of people could potentially be analyzed similarly. For ideas (or memes \cite{Dawkins1989}) to be transmitted through a book (entity, $E$), the book needs to be read by a person (sensor, $S$) and its meaning (property, $P$) understood. A reader who writes with or without changing the ideas in the original book is effectively transmitting information across one `generation'. (Perhaps, the many unread books are akin to the unmeasured molecules that crowd. Such crowding could narrow the focus of the reader on a few books and potentially change the nature of the books that are written.) Collective analysis of many such transmissions through books and other media may provide insights into the origins of a culture or zeitgeist.

The homeostatic preservation of cell codes in successive generations that living systems achieve using their entities, sensors, and properties could inform the design, analysis, and construction of other complex adaptive systems. To apply the Entity-Sensor-Property framework, the information content in a complex system needs to be parsed in terms of these three parameters. Organizations, economies, social networks, ecosystems etc., may all be amenable to such parsing. Indeed, similar sensor-based detection and control has been proposed even for the analysis of human behavior \cite{Powers1960}. In cases where the constituent parts of a system are not known or are unknowable \cite{Rosen2000}, simulations exploring a variety of possible entities, sensors, and properties could help constrain hypotheses.

In summary, the framework developed here for heritable information in living systems is applicable across many scales and therefore may be a generally useful lens for viewing other persistent adapting systems.

\vspace{10pt}
\pagebreak
\noindent{\Large{\textbf{Supplemental Figures and Figure Legends}}}

\vspace{10pt}

\begin{figure*}[h]
  \centering
  \includegraphics[width=0.4\linewidth]{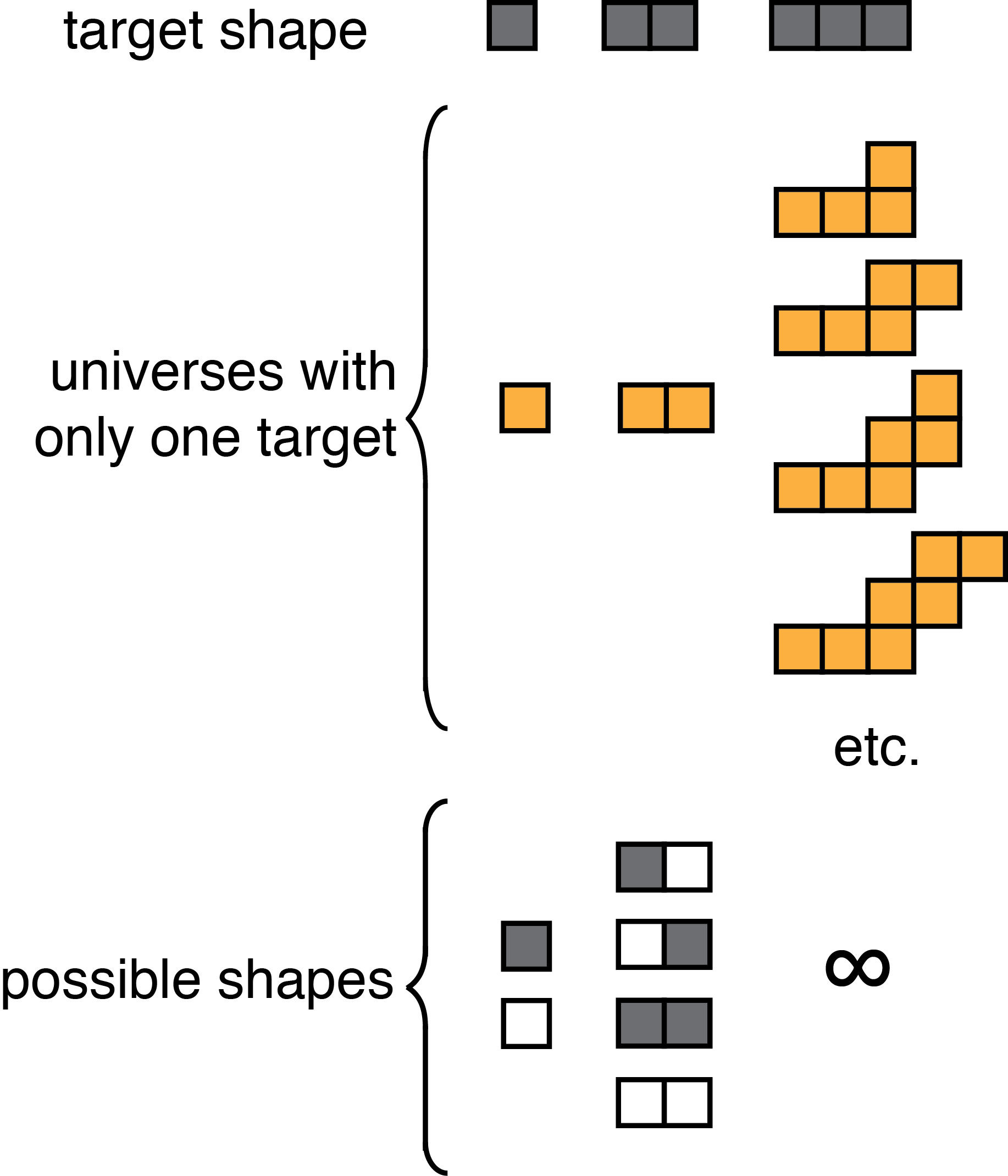}
\caption{\small\textbf{Supplemental Figure 1.} The maximal information content in non-trivial shapes is incalculably large because the number of universes that can be constructed to contain a target shape once and only once is infinite (see supplemental text for details).}
\label{fig:S1}
\end{figure*}
\begin{figure}
  \centering
  \includegraphics[width=0.5\linewidth]{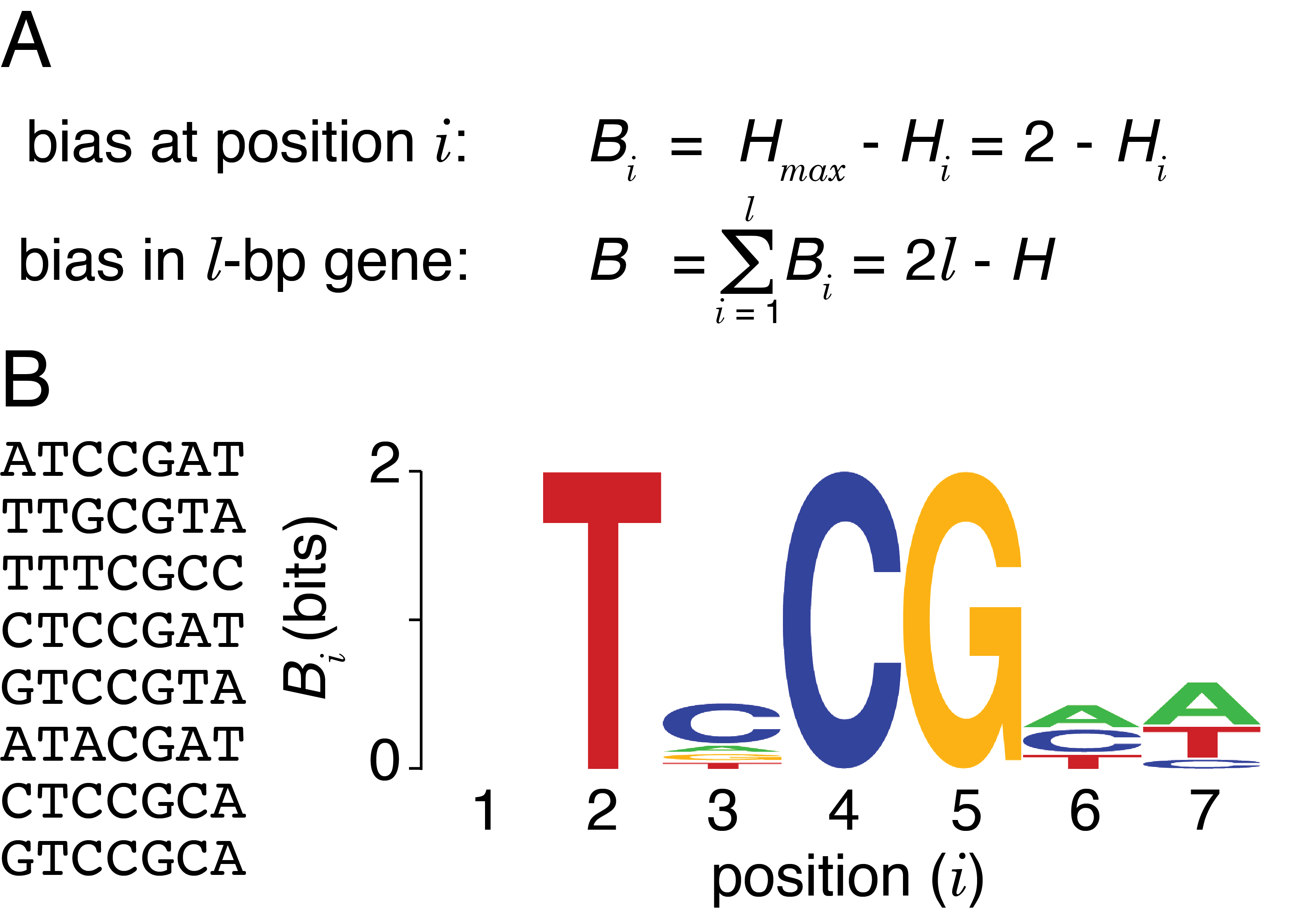}
  \caption{\small\textbf{Supplemental Figure 2.} Conserved bases among a set of sequences reduce the capacity for storing new information in any one sequence. (A) Sequence bias measures the reduction in the capacity to store information. The bias at any position in a 4-base genome is given by $B_i = H_{max} - H_i = 2 - H_i$, where $H_{max}$ and $H_i$ are the maximal and observed storable information at a position $i$. For a gene of length $l$, the total bias is given by $B = \sum_{i}B_i = l \sbullet log_2{4} - \sum_{i}H_i = 2l - H$. (B) Sequence bias reduces available space for storing new information. \textit{Left}, A set of aligned sequences made of 4 bases with varying degrees of conservation at individual positions. \textit{Right}, Bits of bias at each position ($B_i$) in the set of sequences depicted as a sequence logo \cite{Schneider1990} using weblogo \cite{Crooks2004} without small sample correction. Thus, greater the conservation among a set of genomes from different organisms, fewer the bases available for storing \textit{new} information in the genome that distinguishes each organism.}
\label{fig:S2}
\end{figure}
\begin{figure}
  \centering
  \includegraphics[width=\linewidth]{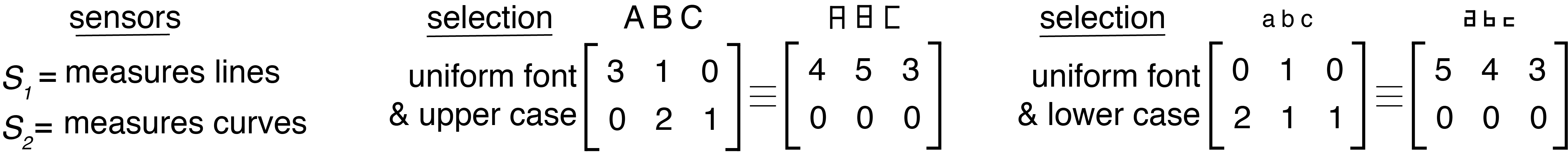}
  \caption{\small\textbf{Supplemental Figure 3.} Distinct internal configurations of the toy model may be seen as equivalent by selection. With the two sensors (left) of the toy model, selection for uniform font \& upper case (middle) or selection for uniform font \& lower case (right) would result in distinct internal states of the cell becoming equivalent. Numbers in matrices are the property values of the contents of the cell in different states ({\fontfamily{phv}{\selectfont ABC}}, {\fontfamily{phv}{\selectfont abc}}, etc.) as measured by the two sensors.}
\label{fig:S3}
\end{figure}
\begin{figure}
  \centering
  \includegraphics[width=0.5\linewidth]{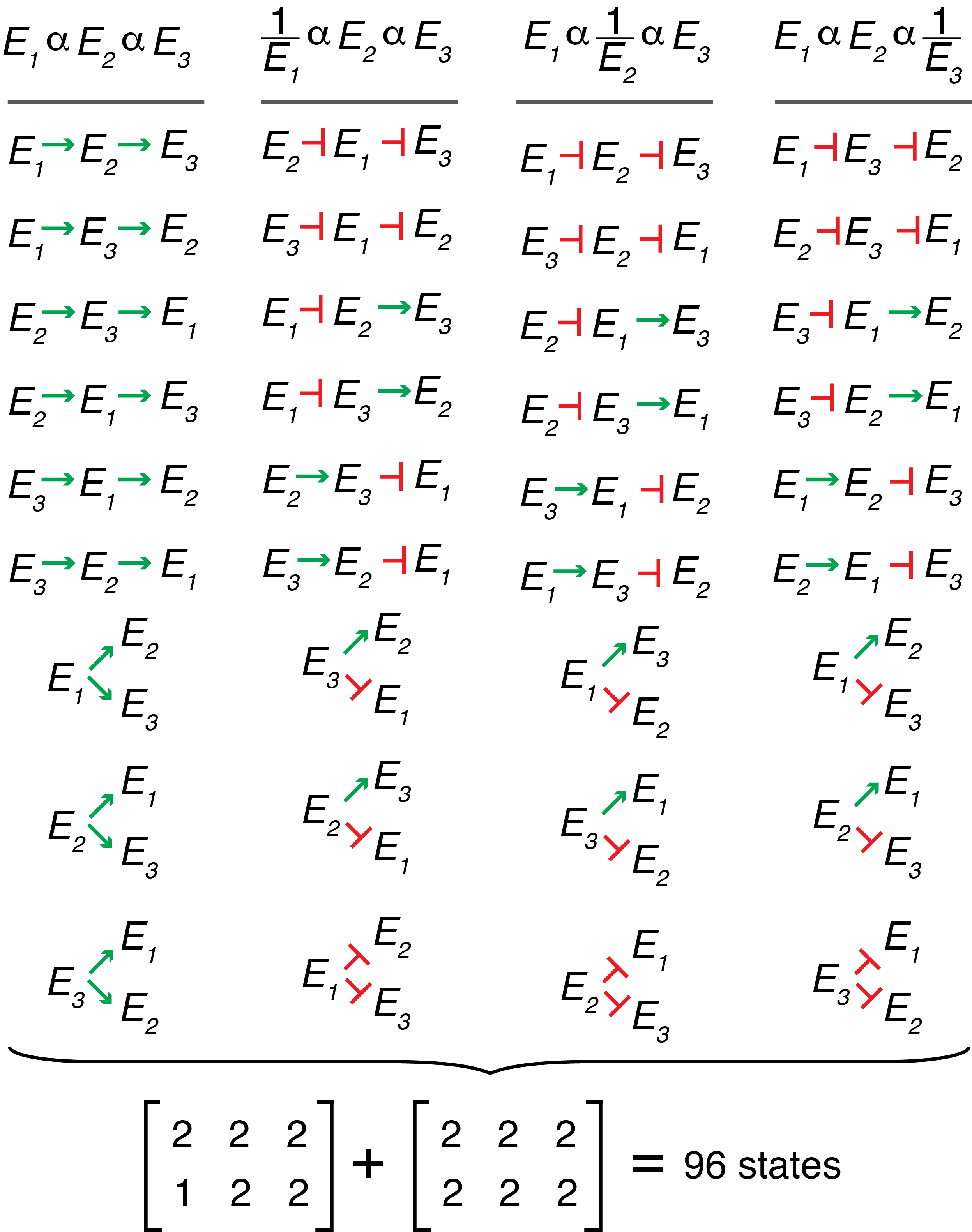}
  \caption{\small\textbf{Supplemental Figure 4.} Impact of two regulatory constraints on the toy model. Different inhibition (bar) or activation (arrow) interactions between all three entities were considered. Four different relationships between the three entities can arise when two interactions are added (9 regulatory architectures each). In all 36 architectures, there are only two sets of values that all entities can take (e.g., when $E_1 \rightarrow E_2 \rightarrow E_3$, then either all entities have the uppercase values or all entities have the lowercase values). This results in the detection of fewer states by each sensor such that every architecture results in the same two matrices of sensed values. Calculating the sum of the products of the elements in each matrix gives a total of 96 states. }
\label{fig:S4}
\end{figure}

\pagebreak

\vspace{10pt}

\noindent{\Large{\textbf{Methods}}}

\vspace{10pt}

Calculations of states for the toy model of an Entity-Sensor-Property system were performed by hand. Distributions of $H$ and $\sigma$ were calculated for one million samples of varying probabilities of base composition using the code below and plotted in R \cite{R2019}.

\begin{lstlisting}[language=R]
S = 0; H = 0; data = list()
for (j in seq(2,10,2)){
for (i in 1:1000000){
vals = matrix(runif(j, 0, 1), nrow = 1, ncol = j)
probs = sapply(vals, FUN = function(x) x/sum(vals))
S[i] = sd(probs)
H[i] = prod(-1, sum(sapply(probs, FUN = function(x) prod(x, log2(x)))))
}
data[[j]] = data.frame(S,H)
}
\end{lstlisting}

For Figure 1, images were captured from simulations created in the Golly application \cite{Golly}. Figure 1A depicts an implementation of von Neumann's self-reproducing universal constructor at the starting stage. This machine is a modification of the original design by von Neumann, but is regarded as the first implementation of his vision of a self-replicating universal constructor. Minor correction of the published design and a modification that reduces the tape length by $\approx$13\% (script by Tim Hutton) were used in this implementation authored by Renato Nobili and Umberto Pesavento. Figure 1B depicts an implementation of the Langton loop. Instructions are stored in a set of dynamic core parts (light blue) that are surrounded by static sheath parts (grey). Each signal packet consists of the signal (yellow, dark blue, orange, magenta, green) followed by a blank (white). This implementation was done by Eli Bachmutsky.
\vspace{10pt}

\noindent{\Large{\textbf{Supplemental References}}}

\printbibliography[heading=none]